\documentclass{article}
\usepackage[utf8]{inputenc}
\usepackage{fullpage}
\usepackage{color}
\usepackage{algorithm,algpseudocode}
\usepackage{graphicx}
\usepackage{amsmath,amssymb}
\usepackage[normalem]{ulem}
\usepackage{authblk}

\title{Group polarization, influence, and domination in online interaction networks: A case study of the 2022 Brazilian elections} 

\author[1]{Ruben Interian}

\author[1]{Francisco A. Rodrigues}

\affil[1]{
Institute of Mathematics and Computer Science, University of São Paulo, São Paulo, Brazil
}

\date{2023}

\begin{document}

\maketitle

\begin{abstract}
In this work, we investigate the evolution of polarization, influence, and domination in online interaction networks. 
Twitter data collected before and during the 2022 Brazilian elections is used as a case study. 
From a theoretical perspective, we develop a methodology called d-modularity that allows discovering the contribution of specific groups to network polarization using the well-known modularity measure. 
While the overall network modularity (somewhat unexpectedly) decreased, the proposed group-oriented approach allows concluding that the contribution of the right-leaning community to this modularity increased, remaining very high during the analyzed period. 
Our methodology is general enough to be used in any situation when the contribution of specific groups to overall network modularity and polarization is needed to investigate. 
Moreover, using the concept of partial domination, we are able to compare the reach of sets of influential profiles from different groups and their ability of accomplishing coordinated communication inside their groups and across segments of the entire network during some specific time window. 
We show that in the whole network the left-leaning high-influential information spreaders dominated, reaching a substantial fraction of users with less spreaders. 
However, when comparing domination inside the groups, the results are inverse. 
Right-leaning spreaders dominate their communities using few nodes, showing as the most capable of accomplishing coordinated communication. 
The results bring evidence of extreme isolation and the ease of accomplishing coordinated communication that characterized right-leaning communities during the 2022 Brazilian elections. 
\end{abstract}

\section{Introduction}

The Global Risks Report published by the World Economic Forum (WEF) is an annual study that explores some of the most severe risks may be faced over the next decade. The authors of the most recent report~\cite{2023wef} state that the erosion of social cohesion and societal polarization ranked as the most significant societal risk among those pointed by WEF 2023. Moreover, societal polarization is listed as the fifth-most severe global risk in the short term (two years) and the seventh-most significant one 
over the next ten years. 
As an example that illustrates these asseverations, an analysis by the Pew Research Center discovered that, on average, Democrats and Republicans are farther apart ideologically today than at any time in the past 50 years in the USA~\cite{2022pew}.

Defined as the fracturing into sharply contrasting communities, polarization leads to declining social stability due to the insufficient communication between polarized, strongly-connected groups~\cite{2021IntMorRib}. 
Extreme polarization, or radicalization, may lead to gridlocks or even violent conflicts~\cite{2015Garcia}. 
Polarization may reduce the space for collective problem-solving, 
stimulating the adoption of short-term policy platforms to galvanize one side of the population~\cite{2023wef}. 
There has been extensive research on the more general issue of competing interactions and the possibility of consensus formation~\cite{2022Weron}. 

Social networks and mass media are places where polarization manifests itself in a strong way~\cite{2018Interian}. 
They have attracted millions of individuals by allowing them to communicate, share their ideas, and discuss different topics, being an excellent tool for studying individual and group behavior. 
However, polarization and hostility are increasingly shifting from social media to the real world, as it was demonstrated by several political events, such as the protests of the Yellow Vest movement in France in 2018, the protests after George Floyd's death in 2020, the US Capitol attack in 2021, and the convoy protests in Canada in 2022~\cite{2022InterianRev}. 

Previous studies analyzed partisan polarization in online participatory platforms, parliaments, and blogging networks in the USA~\cite{2021Dinkelberg,2017Garimella_Long_term,2017Shi}, 
France~\cite{2021RamaciottiMorales}, Italy~\cite{2015DalMaso},   Israel~\cite{2021Wolfowicz}, India and Pakistan~\cite{2020Haq}, and a group of 16 European countries~\cite{2010maoz}. 
Markgraf and Schoch~\cite{2019Markgraf} reported a case study based on data from the German Federal Election of 2017 for illustrating their echo chambers research framework. 

It is known that in the last decade, Brazilian society has been highly polarized. 
Cota et al.~\cite{2019Cota} analyzed the effects of echo chambers in information spreading in a Twitter interaction network related to the impeachment of the former Brazilian President Dilma Rousseff. 
They showed that, on average, users with pro-impeachment leanings could transmit information to a larger audience than users expressing anti-impeachment inclinations. 
Online polarization in Brazil was also analyzed in~\cite{Soares2019}, showing how news outlets influenced political discussions during the 2018 Brazilian presidential campaign. 

In this study, we investigated the evolution of polarization, influence, and domination during the 2022 presidential election in Brazil. 
In 2022, president Lula da Silva won the Brazilian presidential election of 2022 by 1.8 points -- the slimmest margin recorded since redemocratization, which happened between  1974 and 1985. 
To the best of our knowledge, this is the first work that analyzes this period. 
Among other implications, this research could help us to identify the reasons that may lead to the coup events that occurred on 8 January 2023 in Brasília, when protesters attempted to forcefully depose Brazil's democratically elected president 
by breaking into and vandalizing the Supreme Federal Court, the National Congress building, and the Planalto Presidential Mansion in the Three Powers Plaza. 
We are also interested in information spreading and domination before and during the elections, both in the interaction network among individuals and in each community individually. 

In our investigation, we used Twitter social network data. 
Many public figures - politicians, personalities, and researchers - use Twitter frequently to communicate and expose their viewpoints, often using this social network as their official public profile~\cite{2015Kaleel}. 
We built the interaction network using over 15 million tweets published by almost 2 million users. 

In our case study, we are interested in answering these specific research questions: 
\begin{enumerate}
    \item How did the interaction network polarization evolved before and during the Brazilian election, and how did specific groups contribute to overall network polarization? 
    \item How was the information spread before and during the elections, both in the interaction network among all users and in each group individually, and what were the differences in 
    the difficulty of achieving coordinated communication between the groups? 
\end{enumerate}

For answering the first research question, we develop a methodology that allows discovering the contribution of specific groups to network polarization using the well-known modularity measure. 
We found that, while the overall modularity (somewhat unexpectedly) decreased, the contribution of the right-leaning group to this modularity stayed very high during the analyzed period. 
The contribution of other groups were less significant, and more unsystematic. 
We present several pieces of evidence to explain this behavior. 
Our methodology is general enough to be used in any network with several communities, and the contribution of specific groups to overall network polarization 
is needed to investigate. 

For the second research question, the concept of partial domination is applied. 
Domination is used for studying coordinated communication in social networks~\cite{2015Campan,1998Haynes}. 
We show that right-leaning and left-leaning high-influential users have differences in the way they dominate their groups and the whole interaction network. We also compared the domination profiles of these groups to those of different, more neutral communities. 


This work is organized as follows. In the next Section, we describe the data capturing process and the collected dataset. Section 3 presents the theoretical background and our methodology, including the approach to assess the contribution of specific groups to network polarization, and the partial domination concept. The results are discussed in Section 4. Concluding remarks and implications are drawn in the last Section. 

\section{Data collection}

We collected 
a large dataset related to the 2022 Brazilian presidential election from the Twitter social network. 
There is no single hashtag that identifies all tweets related to the election, since the most used hashtags covered only a small proportion of election-related tweets. Therefore, for constructing our dataset, we used a query composed of a list of election-related words in Portuguese: \emph{eleição OR eleições OR eleitoral OR eleitorais}. 
The period for data collection was from September 17 to October 30, 2022, resulting in more than 15 million tweets. 

\begin{figure}
    \centering
    \includegraphics[width=1\textwidth]{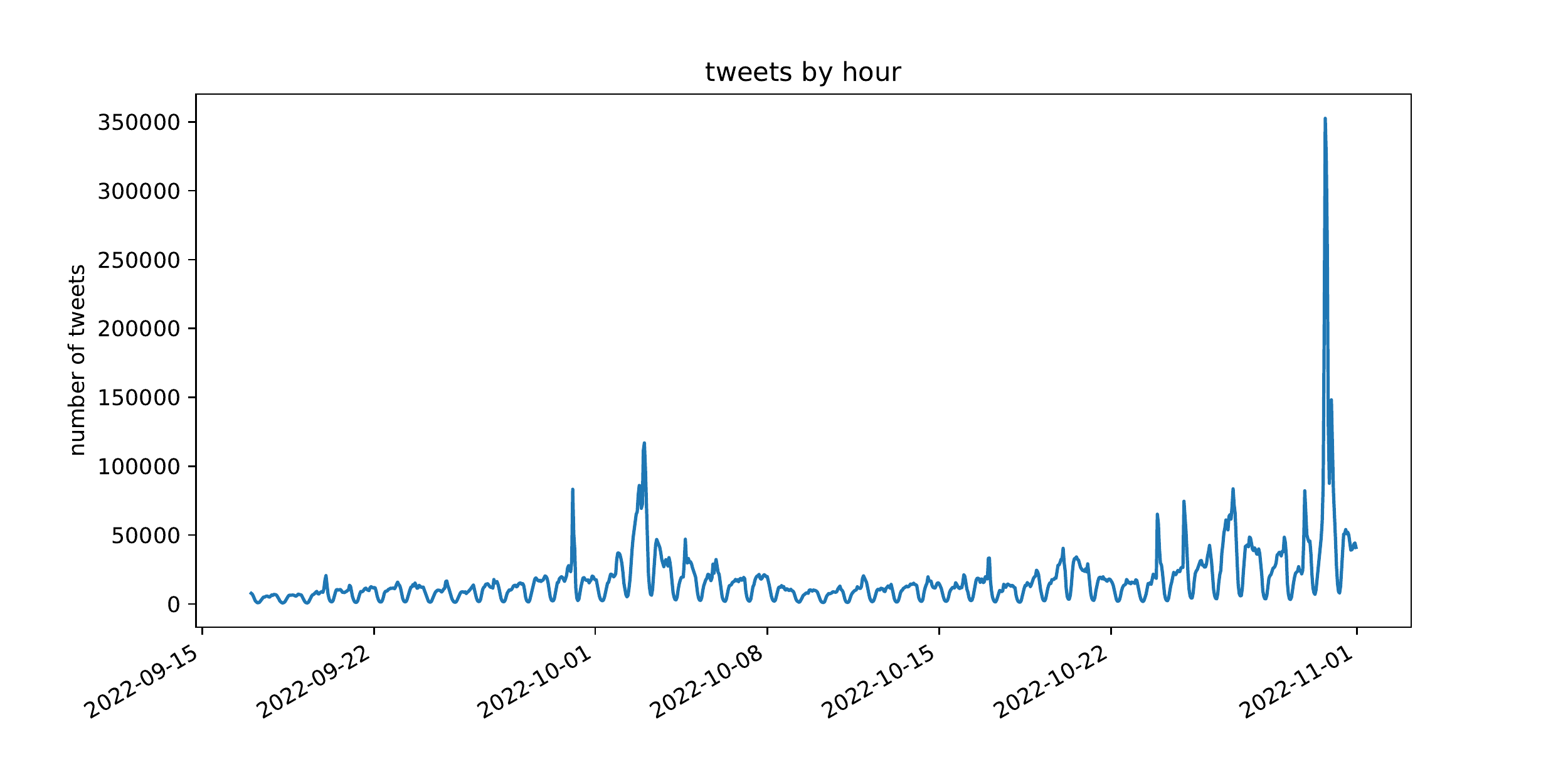}
    \caption{Number of tweets published by hour during the period from September 17 to October 30, 2022. } 
    \label{fig:tweets_by_hour}
\end{figure}

The collected data can be roughly divided in two periods: the period of 15 days before the first round of voting, including the day of the elections first round that took place on October 2, 2022; and the period between the two rounds of elections. 
The second round happened on October 30, 2022. 
Figure~\ref{fig:tweets_by_hour} shows the users' hourly tweeting pattern during the considered period. 
The curve reaches two maximums, corresponding to the days of two rounds of the elections. 

Table~\ref{tab:stats} shows the main characteristics of the collected data, namely the number of collected tweets, unique users, and the number of tweets by its type: retweet, reply, quote. This classification is not a partition, since a tweet can be at the same time a quote and a reply. A tweet may also be unclassified if it is not replying, quoting, or reposting another message. As can be seen, most tweets are retweets, generally 
representing endorsements of others' opinions. 

\begin{table}[ht]
    \centering
    \begin{tabular}{l|r}
    \textbf{Statistic} & \textbf{Value} \\\hline
    Number of tweet authors &  1,890,956 \\
    Number of tweets        & 15,208,839 \\
    Number of retweets      & 13,449,880 \\
    Number of replies       & 1,354,211 \\
    Number of quotes        & 237,601 
    \end{tabular}
    \caption{Main descriptive statistics of the collected Twitter dataset on the 2022 Brazilian election. }
    \label{tab:stats}
\end{table}

\section{Methods}

We have shown that most tweets (specifically, 88.4\%) are retweets, representing some kind of support for the opinions of others. 
These links generally reflect positive relations to the idea or message shared by others. 
For this reason, we used a network approach to model the interactions during the analyzed time window. 

We model the retweet network as a directed graph $G(V, A)$, being $V$ a set of $n$ vertices that represents the users, and $A$ a set of $m$ arcs representing the retweets. An arc $a=(x, y)$ indicates that the user $y$ retweeted one or more tweets published by the user $x$, and indicate the direction of information propagation, from $x$ to $y$. 
We also created networks that represent the interactions restricted to each day. 

\subsection{Community detection} 

Conover et al. showed that the community structure of 
retweet networks can be used to predict the political alignment of users with at least 95\% accuracy~\cite{Conover2011}. 
The graph of retweets analyzed in this work is sparse, and very large, contaning 1,203,164 vertices (Twitter users) and 8,504,687 arcs, each representing one or more retweets. 

Due to the size of our dataset, the choice of the community detection method is quite restricted. 
For these reason, we used the community detection approach based on the multilevel algorithm of Blondel et al.~\cite{Blondel2008}. 
Computer simulations on large networks show that the complexity of this algorithm is linear on sparse graphs~\cite{Blondel2008}. 
The period used by the community detection algorithm corresponded to all tweets published from September 17 to October 29, 2022. 
The election day (October 30, 2022) was excluded because it presented a highly uncommon pattern compared to the rest of the period; thus, excluding it, we guarantee that last-hour binary decisions on the final data collection day will not affect the communities formed during the period. 
However, the data collected during second round of elections was used in the rest of the study. 

The three largest user groups identified by the community detection algorithm are presented in Table~\ref{tab:groups}, ordered by their size. 
They are easily identified with three large digital communities. 
The largest one correspond to left-leaning users. 
The second one correspond mostly to news media and press followers. 
The third one contains right-leaning users. 
Together these three groups cover 86.3\% of the users set. 
There were many other groups whose size was significantly smaller. 
User memberships in these groups were also considered in our analysis, but specific results are not presented for these communities. 

\begin{table}[ht]
    \centering
    \begin{tabular}{l|c|r}
    \textbf{Community} & \textbf{\# of users} & \textbf{Some representative members} \\\hline
    Left-leaning & 552,578 & @LulaOficial, @siteptbr, @HaddadDebochado \\
    News media, press & 244,296 & @JornalOGlobo, @g1, @UOLNoticias, @folha \\
    Right-leaning & 241,641 & @JovemPanNews, @revistaoeste, @jairbolsonaro 
    \end{tabular}
    \caption{Largest communities identified by the multilevel algorithm by Blondel et al.~\cite{Blondel2008}. }
    \label{tab:groups}
\end{table}

\subsection{Modularity and contribution of groups to polarization}

For a given division of the network's vertices into some collection of groups or communities, modularity~\cite{2004Newman} reflects the concentration of edges within groups compared with random distribution of edges between all vertices regardless of the groups. 
We use the modularity of the underlying undirected graph of $G(V, A)$ for measuring the polarization of the network we built with the collection of communities obtained in the previous section. 

More formally, let's suppose that a graph $G(V, E)$ with $|V|=n$ vertices and $|E|=m$ edges, have its vertex set $V$ partitioned into $k$ disjoint groups $\{A_1, A_2, \ldots, A_k\}$. The modularity $Q$ is defined as: 

\begin{equation}
Q = \frac{1}{2m} \sum_{u,v \in V}(a_{uv} - \frac{d(u) d(v)}{2m}) \cdot \delta_{g_u g_v}, 
\label{eqn:modularity}
\end{equation}
\noindent
where $d(v)$ is the degree of node $v \in V$; 
$g_v$ the index of $v$'s group; 
$a_{uv} = 1$ if there is an edge between nodes $u$ and $v$, $0$ otherwise; 
and $\delta_{ij}$ is the Kronecker delta. 

While modularity can measure the overall level of polarization of the retweet network for some time window, it is not designed to evaluate the contribution of each group to network's polarization. 
Note that if we evaluate the modularity of the subgraph induced by some group $A_i$ (i.e., a subgraph that contains all vertices in $A_i$ and all edges between vertices in $A_i$), then the modularity value $Q$ for this single group will be equal to zero~\cite{2016Newman}. A previous study showed that there is a lack of consolidated group-level measures for evaluating polarization~\cite{2022InterianRev}. 

For evaluating the contribution of some specific group to the overall network modularity, we developed an approach called $d-modularity$. 
Since $\delta_{g_u g_v} = 1$ if and only if the vertices $u$ and $v$ are from the same group (and is zero otherwise), we can rewrite the equation~(\ref{eqn:modularity}) that defines modularity $Q$ as follows: 

\begin{equation}
Q = \frac{1}{2m} \sum_{i=1}^{k} \; \sum_{u,v \in A_i}(a_{uv} - \frac{d(u) d(v)}{2m})
\label{eqn:modularity_variant}
\end{equation}

Now let's suppose that one specific group $A_i$ is chosen. 
Let us consider one addend in equation~(\ref{eqn:modularity_variant}) that corresponds to the group $A_i$ (the same reasoning applies to each group $A_i = A_1, A_2, \ldots, A_k$). 
Note that this term $\sum_{u,v \in A_i}(a_{uv} - d(u) d(v) / 2m)$ considers only pairs of vertices $u$ and $v$ that both belong to $A_i$, and $\sum a_{uv}$ 
represents the actual number of in-group edges. 
Moreover, $\sum d(u) d(v) / 2m$ 
represents the expected number of $A_i$'s in-group edges after rewiring or randomizing the edges in the network while preserving the degree of every vertex (randomization known as the configuration model). 
If group $A_i$ gives no more within-community edges than would be expected by random chance, then the whole term $\sum_{u,v \in A_i}(a_{uv} - d(u) d(v) / 2m)$ will be nullified. On the other hand, if group $A_i$ gives significantly more within-community edges than would be expected by random chance, then the contribution of this group to network modularity will be large. 

Therefore, we propose the following way to measure the contribution of a specific group $A_i$ to the overall modularity $Q$ of the network. 
Let the contribution $Q_i$ of the group $A_i$ to network's modularity be defined in the following way: 

$$
Q_i = \frac{1}{2m} \sum_{u,v \in A_i}(a_{uv} - \frac{d(u) d(v)}{2m})
$$

Note that, as expected, $\sum_{i=1}^{k} Q_i = Q$, i.e., the sum of the contributions of all groups produces the overall network modularity. 
Now, the $d-$modularity $d_i$ of $G$ respect to the group $A_i$ is defined as: 

$$
d_i = \frac{Q_i}{Q},
$$




In short, the $d-$modularity $d_i$ reflects the relative contribution of some specific group $A_i$ to the overall modularity of the network. 

\begin{figure}
    \centering
    \includegraphics[width=0.45\textwidth]{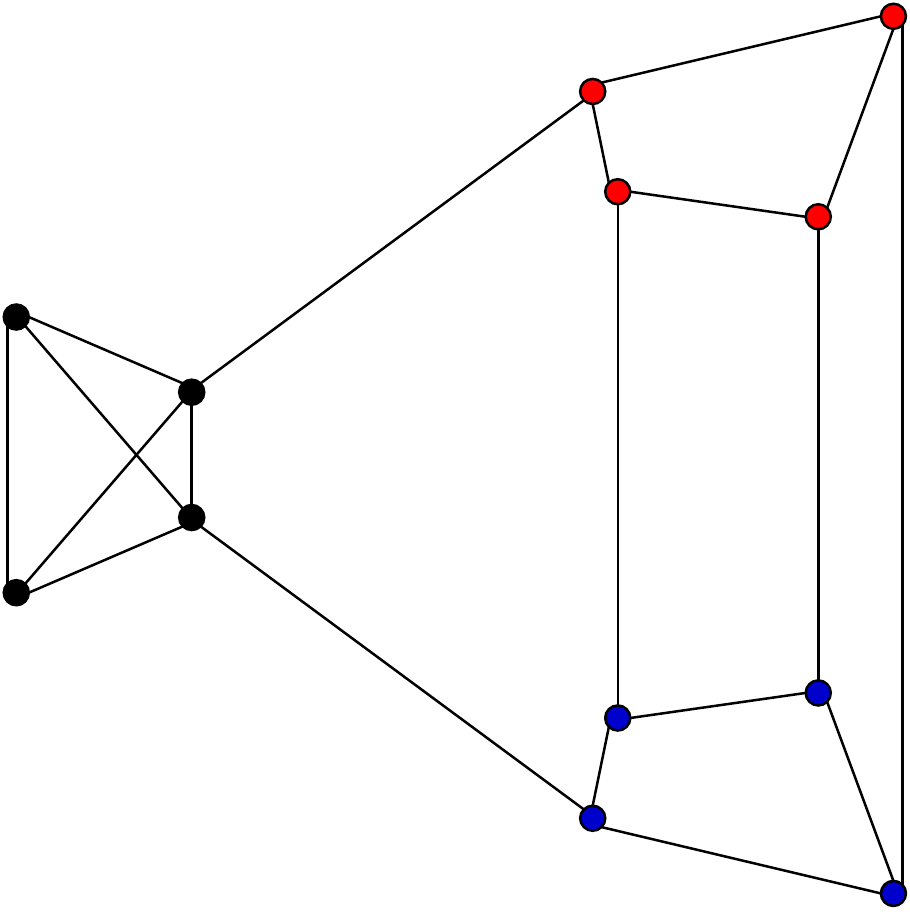}
    \caption{Example of a graph with three groups that contribute unequally to network's modularity $Q = 0.402$. The contribution $Q_1$ of the black group is $0.180$ ($d_1=0.448$, or $44.8\%$), while the contribution of red and blue groups are $Q_2=Q_3=0.111$ ($d_2=d_3=0.276$, or $27.6\%$). } 
    \label{fig:example_d_mod}
\end{figure}

Figure~\ref{fig:example_d_mod} shows an example of a graph with 12 vertices partitioned into three groups, whose overall modularity is $0.402$. 
The three communities are equally-sized, with four vertices each. 
Note that each red or blue vertex has exactly two in-group adjacent vertices, and one adjacent vertex from a different group. 
Each black vertex, however, has exactly three in-group neighbors, and at most one adjacent vertex from a different group. 
Correspondingly, the group $A_1$ composed of black nodes contributes with $Q_1 = 0.180$ to modularity, thus $d_1 = 0.448$, or 44.8\%. On the other hand, the contribution of the other two groups $A_2$ and $A_3$, composed of red and blue vertices, respectively, is quite smaller ($Q_2=Q_3=0.111$, and $d_2=d_3=0.276$, or $27.6\%$). 

To illustrate other possible $d-modularity$ values that might occur in practice, suppose that we have a graph $G(V, E)$ with strong community structure (modularity values in the range from 0.3 to 0.7~\cite{2004Newman}), several groups, and one specific group $A_i$ is chosen. 
We highlight four possibilities regarding the $d-$modularity value $d_i$ obtained using the described procedure: 

\vspace{5mm}
\begin{enumerate}
    \item 
    If $d_i \approx 1$, then $Q_i \approx Q$, and the network's modularity is strongly affected by the connections of vertices in $A_i$. 
    This fact indicates that the contribution of $A_i$ to the polarization of the network ($A_i$'s distribution of within-group and out-group edges) is decisive for overall network's strong community structure. 
    
    \item If $d_i > 0$ (but is neither too close to 1 nor too close to 0), then the contribution of $A_i$ to the polarization of the network is not negligible. Still, neither is it the only one that affects the network's modularity. 
    
    \item If $d_i \approx 0$, then $Q_i \approx 0$, and the network's modularity is not affected significantly by $A_i$'s vertices connections. 
    In this case, the chosen group's contribution to network polarization is neutral, and the distribution of within-group and out-group edges in $A_i$ does not differ too much from that expected by random chance, implying that the strong community structure in the graph is mostly influenced by the contributions $Q_j$ of the rest of groups, with $j = 1, \ldots, k, j \neq i$. 
    
    \item If $d_i < 0$, then the contribution of $A_i$ to the network's modularity is negative. This case is improbable to happen in real networks, neither occurred in our case study, since it implies heterophily in the group's vertex connections. 
\end{enumerate}

The analysis of the retweet network's modularity and the $d-$modularity values was performed day-by-day, that is, considering, at each step, the interactions performed during each 24-hour period. 
The goal was to measure the evolution of network polarization during the analyzed period, as well as the contribution of groups to overall Twitter network polarization. 


\subsection{Domination in interaction graphs}
\label{sec:domination}

In this section, we are interested in studying coordinated information spread both in the interaction network among all users and in each group separately, as well as differences in information spread among the groups. 
For this purpose, we used the concept of domination. 
The mathematical study of domination in graphs comes from the 1960s. 
More recently, the use of domination for the study of coordinated communication in social networks was studied by Campan et al.~\cite{2015Campan}. 
The reader is referred to the book by Haynes et al.~\cite{1998Haynes} for a more detailed review of domination in graphs. 

In this work we intend to find some (expected to be small) subset of information spreaders that is capable of reaching (\textit{dominating}) all or most of the users that participate in a discussion. 
Our research hypothesis is that in real life, small subsets of information spreaders may be used for the task of reaching (or communicating with) a large number of users in some network or in some specific group. 

More formally, the problem can be stated in the following way. Let $V$ be the set of $n$ users, and $D \subseteq V$ be the set of spreaders, i.e., users whose tweets were shared or retweeted at least once. 
We are interested in finding the smallest subset $S^*$ of $D$ such that the number of users reached from $S^*$ is at least $\rho \cdot n$, where $\rho$ is a parameter that reflects the minimum proportion of users to be reached. 
We may study this problem considering, for example, $\rho=$ 100\%, $\rho=$ 75\%, or $\rho=$ 50\% of a given network or group. 
A variant of this problem (which will not be analyzed in this section) proposes finding the smallest subset $S^*$ not of $D$, but of some specific subset $D'$, representing a specific category of spreaders. 

In the case when $\rho=$ 100\%, the abovementioned problem can be formulated as the Directed Dominating Set problem (DDS) previously studied, for example, by Habibulla et al.~\cite{2015Habibulla}. 
A directed graph $G(V,A)$ is built, where $V$ is the set of vertices that represent the users, and $A$ is the set of arcs representing relations between the users. 
The smallest subset $S^* \subseteq V$ that reaches each vertex represents a group of spreaders that dominates the interaction network. 
An instance of the classical Dominating Set problem (DS), known to be NP-hard, can be easily reduced to an instance of DDS problem by replacing each undirected edge $(u, v)$ with a pair of directed arcs $\overrightarrow{uv}$ and $\overrightarrow{vu}$. Consequently, DDS is also NP-hard, that is, there is no efficient algorithm for finding the optimum solution of this problem unless $P=NP$. 

In the more general case when $\rho \leq$ 100\%, however, the problem requires a specific formulation, which we call Partial Directed Dominating Set problem (PDDS). 
An instance of the PDDS problem is composed of a directed graph $G(V,A)$ and a parameter $\rho$. 
The goal is to find a dominating set that reaches at least ${\rho}|V| = {\rho}n$ vertices. 
This problem is similar to the variant of the Set Cover problem called Partial Cover~\cite{1990Kearns}. 
The PDDS is also NP-hard, since it contains the DDS problem as a special case ($\rho = 1$). 

We note that there is also another common formulation for problems involving partial covering. 
In this variant, instead of $\rho$, there is an integer $k$, such as the number of users reached from $S^*$ must be at least $k$. It is easy to show that this versions are equivalent to our formulation by setting $k = \lceil \rho \cdot n \rceil$. 

Since the problem is NP-hard, and the considered instances are very large, a greedy algorithm is used. 
Algorithm~\ref{alg:heuristic} presents the greedy heuristic for the PDDS problem. 
The algorithm is an adaptation of an existing greedy heuristic for the Dominating Set problem studied, for example, by Parekh~\cite{1991Parekh}. It is known to be the best in terms of the approximation ratio for Dominating Set unless $P = NP$~\cite{2004Chlebik}. 
In the algorithm, $W(v)$ denotes the set of yet uncovered vertices among successors of the vertex $v$, also called the span of $v$. 

The algorithm starts setting the current solution $S^*$ to an empty set (line 1), and the number of covered vertices $f^*$ to zero (line 2). 
Initially, the span of each vertex $v$ is initialized by the set of their out-neighbors $\mathcal{N^+}(v)$ (line 3). 
In this notation, the set $\mathcal{N^+}(v)$ includes $v$. 
The algorithm performs a number of iterations (lines 4-9). 
In each iteration, the algorithm picks the vertex with the largest span, i.e., the spreader that reaches the maximum number of uncovered users, updating the current solution $S^*$, the number of covered vertices $f^*$, and the span $W(v)$. 
When the number of covered vertices $f^*$ reaches at least $\rho |V|$ vertices, the algorithm ends returning $S^*$ (line 10).

\begin{algorithm}[ht]
\caption{Greedy algorithm for Partial Directed Dominating Set problem}
\label{alg:heuristic}
\hspace*{\algorithmicindent} \textbf{Input: $G=(V, A)$, constant $\rho$}\\
\hspace*{\algorithmicindent} \textbf{Output: $S^*$}
\begin{algorithmic}[1]
\State $S^* \gets \emptyset$; 
\State $f^* = 0$; 
\State $W(v) \gets \mathcal{N^+}(v) \quad \forall\ v \in V$ 
\While{$ f^* < \rho |V| $}
    \State $u \gets \textit{argmax}_{v \in V} |W(v)| $;
    \State $S^* \gets S^* \cup \{u\}$; 
    \State $f^* = f^* + |W(u)|$; 
    \State $W(v) \gets W(v) \setminus W(u) \quad \forall\ v \in V$
\EndWhile;
\State \textbf{return} $S^*$;
\end{algorithmic}
\end{algorithm}

Note that, since any instance of the Dominating Set problem in undirected graphs can be linearly transformed in a PDDS instance with $\rho=1$ by replacing each undirected edge $(u, v)$ with a pair of directed arcs ($\overrightarrow{uv}$, $\overrightarrow{vu}$), it can be easily verified that the solutions are in a one-to-one correspondence. 
Therefore, as shown in~\cite{1991Parekh}, the greedy algorithm (nor any other one) cannot do better than $\mathcal{H}(\delta + 1)$ times the size of an optimal dominating set, where $\delta$ is the maximum out-degree in the graph, and 
$
\mathcal{H}(\delta + 1) = \sum_{i=1}^{\delta + 1} 1 / i = \Theta(\log \delta) 
$, and
there is no $o(\log \delta)-$approximation for this problem unless $P = NP$. 

In this work, several 
parameters are explored when building instances for the Partial Directed Dominating Set from our data. 
Different values of $\rho$ are used. 
Moreover, we study the domination in the subgraphs that represent the interactions inside user groups, as well as in the whole network. 






\section{Results}

This section presents the main findings of our study, which provide several insights into the evolution of polarization, influence, and domination during the 2022 Brazilian elections. 
The results bring evidence of the extreme isolation of specific communities and the ease of accomplishing coordinated communication inside these groups. 

\begin{figure}
\centering
\includegraphics[width=1\textwidth]{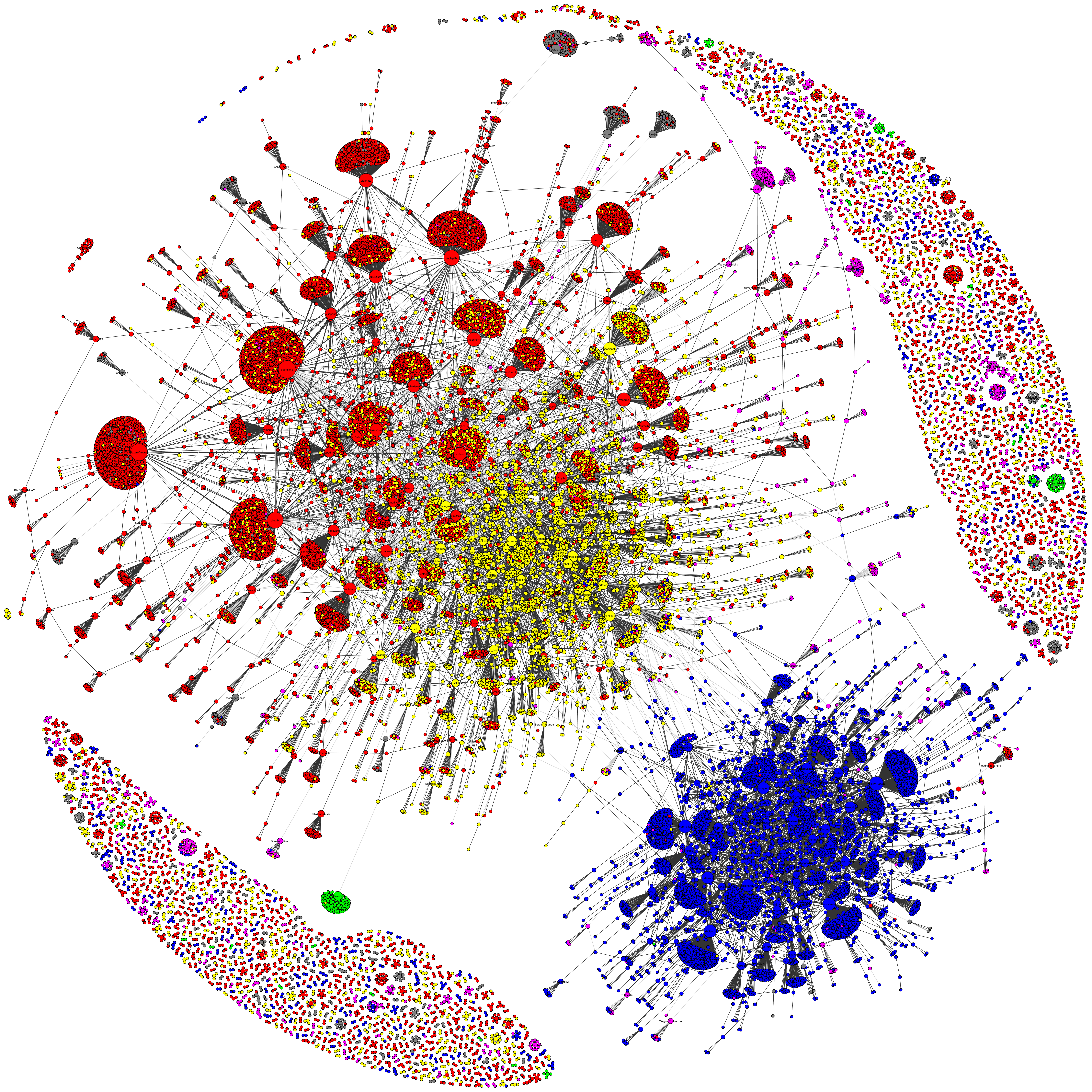}
\caption{Network that represents the interactions during October 2, 2022, the day of the first round of the elections. Red, blue, and yellow vertex colors indicate membership in the left-leaning community, right-leaning community, and media followers group, respectively. } 
\label{fig:graph_election_day}
\end{figure}

Figure~\ref{fig:graph_election_day} exemplifies the generated interaction networks, presenting the graph of interactions that occurred on October 2, 2022, the day of the first round of the elections. 
We recall that the communities represented by red, blue, and yellow vertex colors, among others, were generated considering all interactions over the analyzed period, not only those that happened during this specific day. 
These colors indicate membership in the left-leaning, right-leaning, and news media or press followers communities, respectively. 
The visualization of the network was generated using only 5\% of randomly chosen user-user interactions among the almost 700,000 generated on October 2. 

\subsection{Polarization dynamics and isolation of groups}

The daily analysis of the retweet network's modularity and $d-$modularity values is presented in this section. 
Our goal is to measure the evolution of network polarization, as well as the contribution of groups to overall polarization of the network of retweets, during the analyzed period. 
We observe that, in practice, modularity values for networks with strong community structure typically fall in the range from about 0.3 to 0.7, and higher values are rare~\cite{2004Newman}. 

\begin{figure}
\centering
\includegraphics[width=1\textwidth]{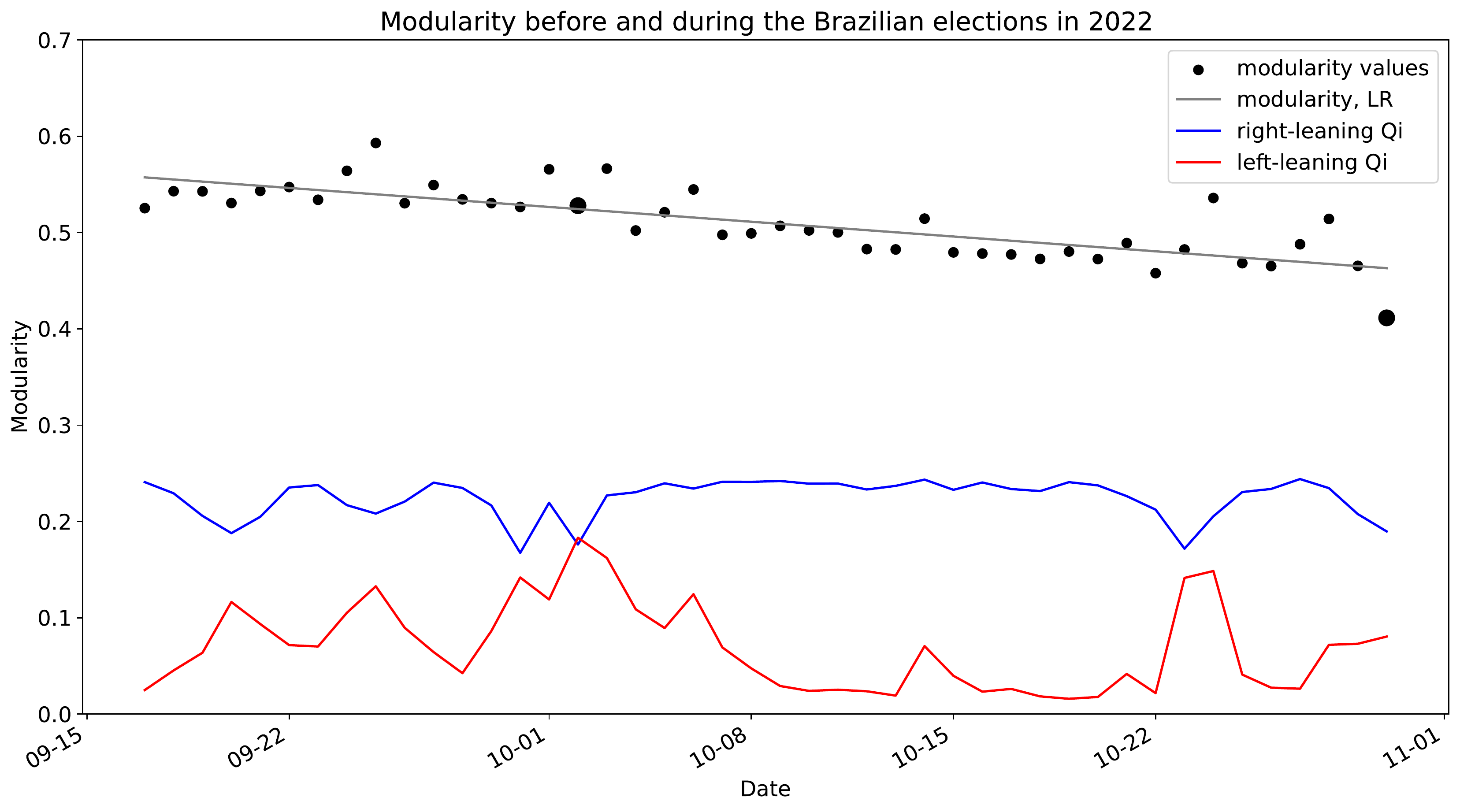}
\caption{Evolution of polarization and group isolation before and during the Brazilian elections in 2022. 
Black points represent modularity values for each day, with two bigger points representing the two rounds of voting days. 
Decreasing gray line shows the modularity trend obtained using least square regression (LR). 
Blue and red lines represent the contributions $Q_i$ of, respectively, right-leaning and left-leaning groups, to network's modularity (see Section 3.2). } 
\label{fig:modularity}
\end{figure}

Figure~\ref{fig:modularity} shows the evolution of polarization and group isolation before and during the Brazilian elections in 2022. 
Despite identifying an unexpected decreasing modularity trend in the whole network, when analyzing the contribution of specific groups, we see a different picture. 
While the overall modularity decreased due to more frequent interactions between users from different groups, 
right-leaning group contribution $Q_r$ to the network's modularity stayed stably high during the entire analyzed period. 
Left-leaning group reached the higher value of $Q_l$ during the first round of elections day, slightly surpassing the right-leaning group contribution $Q_r$, also getting very close to right-leaning group's curve between October 23 and 24, a week before the second round. 
However, on all other analyzed days, right-leaning group's contribution $Q_r$ was higher. 
The curves meet each other only near the first round of elections (on October 2, 2022). 

Note that the right-leaning group is significantly smaller than the group of left-leaning users (see Table~\ref{tab:groups}). 
However, it's contribution to network's modularity is higher than the contribution of the left-leaning, significantly larger, group. 


\begin{figure}
\centering
\includegraphics[width=1\textwidth]{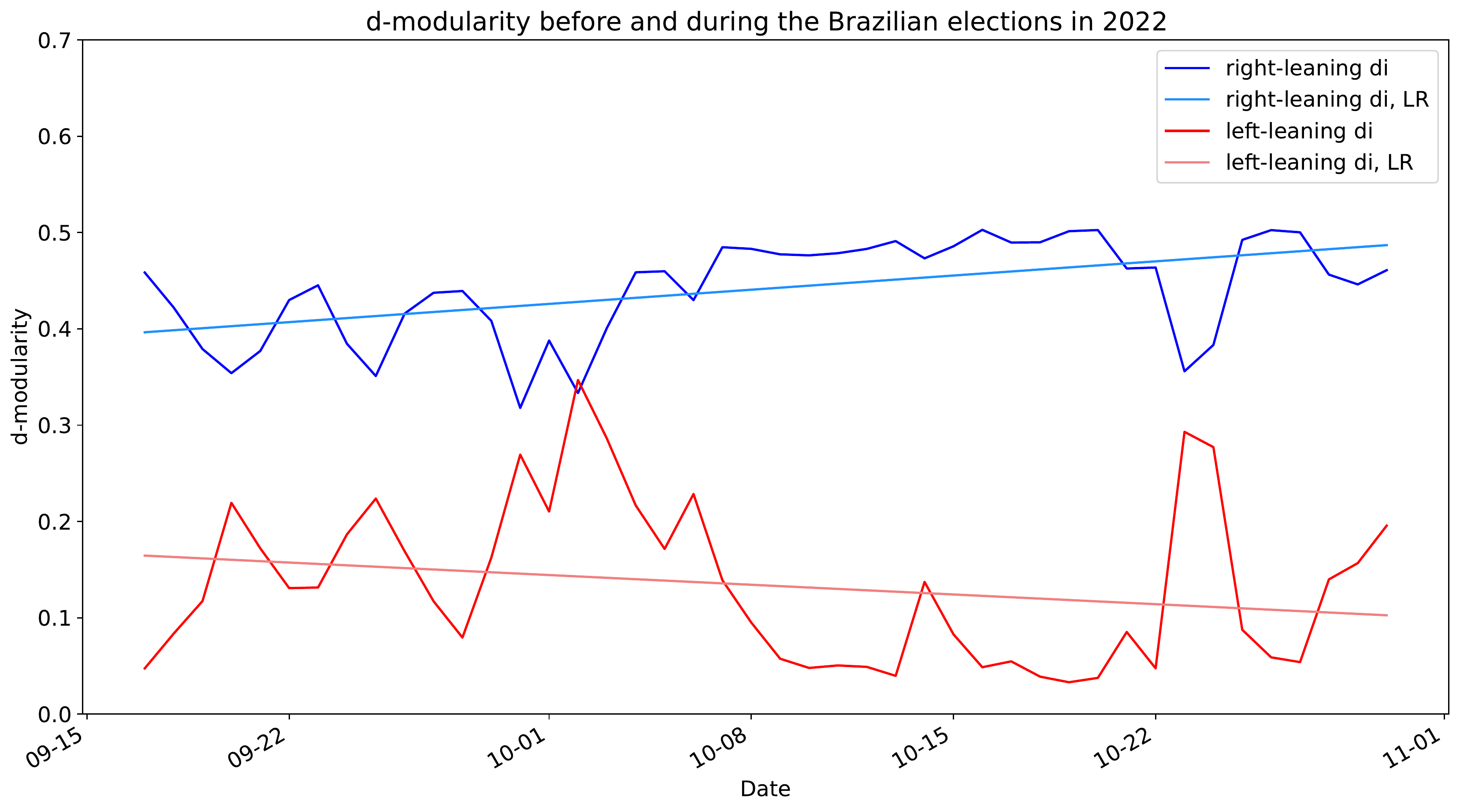}
\caption{Evolution of $d-$modularity before and during the Brazilian elections in 2022. 
Blue and red lines represent the $d-$modularity values $d_i$ for, respectively, right-leaning and left-leaning groups. 
Light blue and light red lines show the $d-$modularity trends obtained using least square regression (LR) over right-leaning and left-leaning $d-$modularity values, respectively. } 
\label{fig:modularity_di}
\end{figure}

On the other hand, Figure~\ref{fig:modularity_di} shows the evolution of $d-$modularity values for right-leaning and left-leaning groups. 
Since there was a stable and high modularity over whole the period ($0.510$ on average, with standard deviation of only $0.036$), the shape is very similar to that of Figure~\ref{fig:modularity}. 
The increasing right-leaning $d-$modularity trend shows that, despite the overall decreasing modularity, the right-leaning group's polarization stayed high and even increased its degree of isolation from other groups. 

On average, during all the analyzed period of 44 days, 44.1\% of the modularity value is explained by right-leaning group's contribution (standard deviation 5.1\%), while the contribution of left-leaning larger group is 13.4\% on average (standard deviation 8.4\%). 
We can conclude that the network's modularity is much less affected by the connections of left-leaning users when compared to the right-leaning ones, since the left-leaning group is much more integrated with the other groups in the network. 

\subsection{Dominating the interaction network during the 2022 Brazilian election}

For measuring domination during the 2022 Brazilian election, we consider two classes of instances. 
In the first class, the goal is to cover vertices in a single group using spreaders from this group alone. 
That is, the instance of the Partial Directed Dominating Set problem (PDDS) is a subgraph whose vertices belong to one single community. 
In this case, we measure the capability of reaching all users in the considered group, as well as the ability to communicate quickly within the group. 
For example, if a small subset of spreaders can reach the majority of the group, the ability of quick communication inside the group is high. 

In the second class of instances, the goal is to cover all vertices in the network by using spreaders from one single group. 
Therefore, the capability of some subset of information spreaders (e.g., right-leaning or left-leaning ones) to reach a substantial segment of users that participate in the discussion is measured. 
Note that in this case a slight modification of the PDDS problem is used, and we attempt to find the smallest subset $S^*$ of some specific subset of spreaders $A_i$. 

\begin{figure}
\centering
\includegraphics[width=0.75\textwidth]{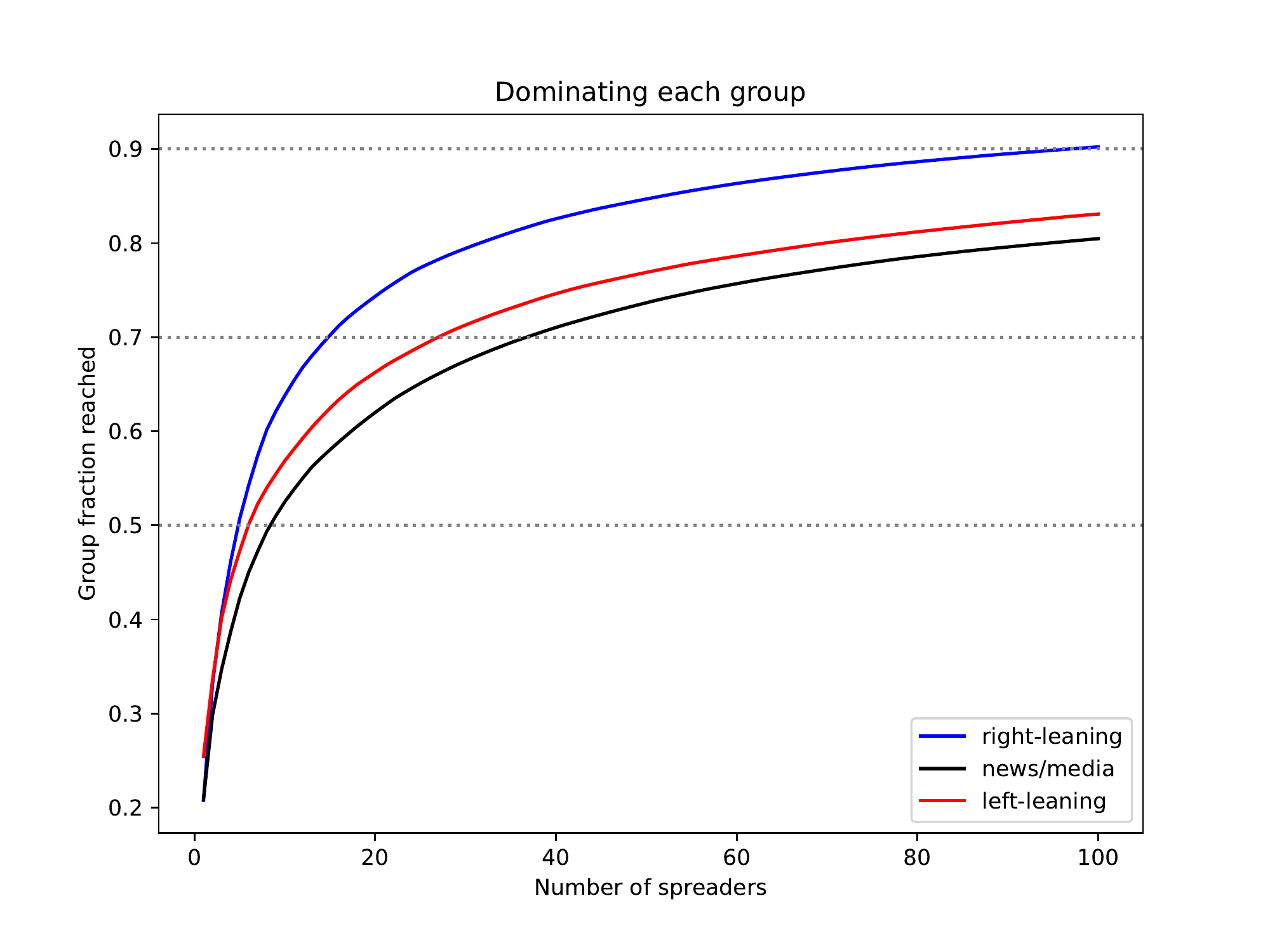}
\caption{Results for in-group domination. Most of the group can be reached using a small group of spreaders. 
For example, the number of spreaders necessary for reaching 70\% of the group are 15, 27, and 37 for right-leaning, left-leaning, and media groups, respectively. 
Right-leaning spreaders are able to dominate up to 90\% of their group using few spreaders. } 
\label{fig:covergroup}
\end{figure}

\begin{figure}
\centering
\includegraphics[width=0.75\textwidth]{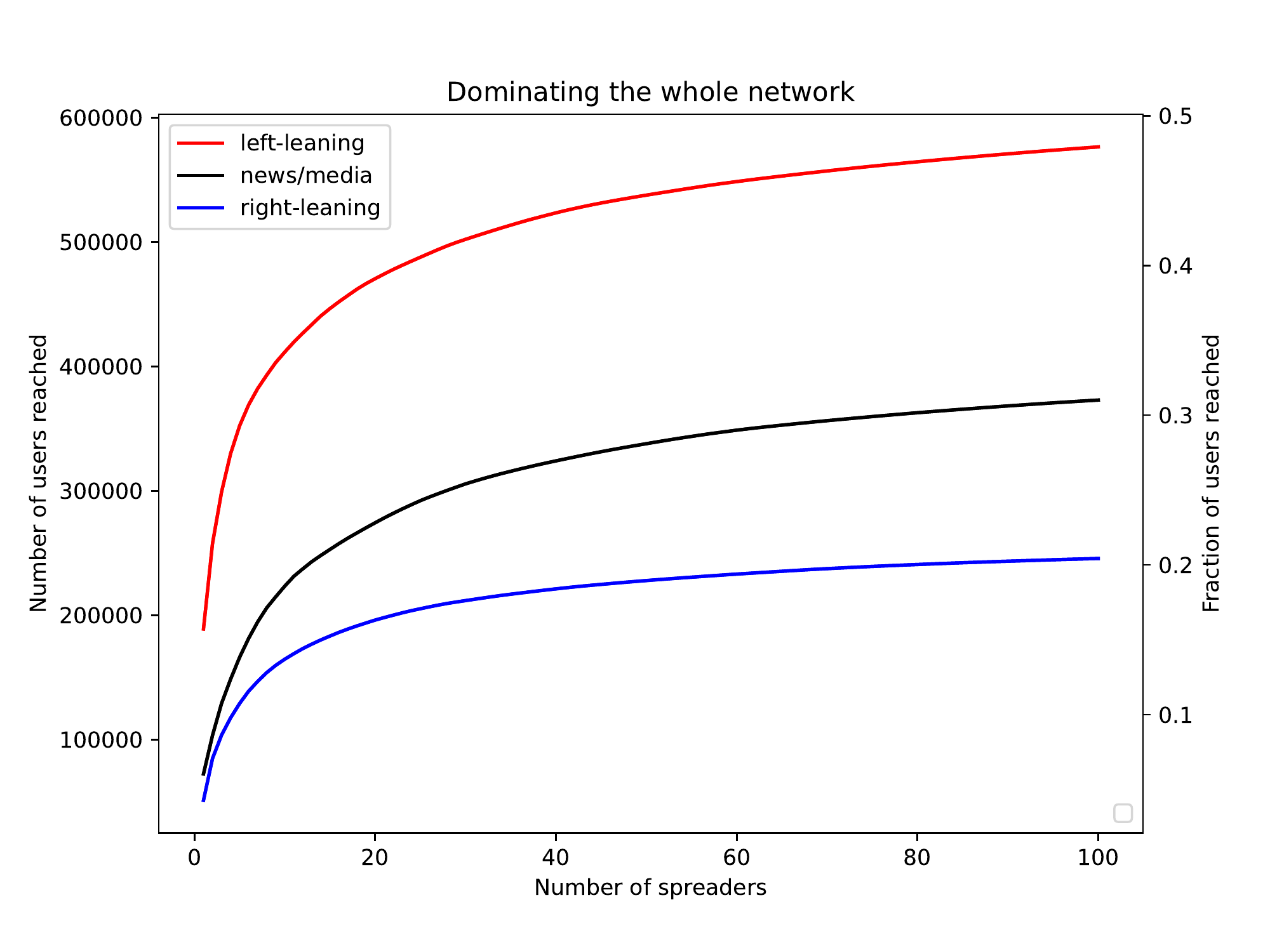}
\caption{Network domination results. Left-leaning spreaders are able to reach a large portion of the network, while right-leaning spreaders may reach a small portion of the network, comparable to the size of their group. } 
\label{fig:covernetwork}
\end{figure}

Figure~\ref{fig:covergroup} 
shows the results for the first class of instances, where the goal is to cover vertices in a single group by spreaders from this group.  
Figure~\ref{fig:covernetwork} 
shows the results for the second class of instances, where the goal is to cover the whole network by using spreaders from one single group. 
We present the results for the left-leaning, right-leaning, and also for the media followers group, taken in order to obtain the domination profile of a more neutral in the ideological spectrum community. 
In each case, the presence of a point $(x_0, \rho_0)$ in the graph means that the greedy Algorithm~\ref{alg:heuristic} described in Section~\ref{sec:domination} returned $x_0$ spreaders in the directed dominating set with algorithm's parameter $\rho$ equal to $\rho_0$. 

Figure~\ref{fig:covernetwork} shows that in the whole network the left-leaning high-influential information spreaders dominated, reaching a substantial fraction of the vertices with less spreaders. 
This outcome is, in a way, expected when considering the 2022 Brazilian election results. 
However, the magnitude of the difference between the groups is very large. 
For example, using at most 100 spreaders, media and right-leaning influential accounts can reach approximately $30\%$ and $20\%$ of users, respectively, very far from the almost $50\%$ reached by 100 left-leaning spreaders. 
This confirms that right-leaning spreaders are confined to their group, which represents approximately 20\% of the users. 

However, when comparing domination inside the groups, we surprisingly found that the results are essentially the inverse. 
As Figure~\ref{fig:covergroup} shows, right-leaning high-influential users dominate their communities using much less vertices than in the other analyzed groups. 
To perceive it, we can draw an horizontal line at, for example, $\rho = 0.7$, and compare the number of spreaders necessary for reaching 70\% of the group obtained by intercepting the curves. 
The obtained values are 15, 27, and 37 for right-leaning, left-leaning, and media groups, respectively. 
Therefore, the differences among the groups are not negligible, and the right-leaning spreaders show as the most capable of accomplishing coordinated communication inside their group. 

A curious but revealing fact is that the most influential tweet from our dataset was made by Casimiro~\cite{2022case}, a non-political Brazilian journalist, presenter, sports commentator, and streamer. As shown in Table~\ref{tab:highest_retweets}, it reached 189,181 retweets in the analyzed period, more than five times the second most influential tweet. 

\begin{table}[hb]
    \centering
    \begin{tabular}{l|c|c|r}
        \textbf{Author's username} & 
        \textbf{Tweet ID} & 
        \textbf{Date} & 
        \textbf{\# of retweets} \\\hline
        @Casimiro   & 1584293297841111040 & October 23, 2022 & 189,181 \\
        @g\_agurgel & 1586727126656602114 & October 30, 2022 & 33,238 \\
        @odontinho  & 1576601689523560449 & October 2, 2022  & 31,306 \\
        @SF\_Moro   & 1577313831449141249 & October 4, 2022  & 29,728 \\
        @RomeuZema & 1577322128231350273 & October 4, 2022  & 28,862 
    \end{tabular}
    \caption{Tweets with the overall highest number of retweets in the analyzed period. }
    \label{tab:highest_retweets}
\end{table} 


\section{Conclusions and implications}

In this study, we investigated the evolution of network and group polarization, influence, and domination in online interaction networks, using Twitter data collected before and during the 2022 Brazilian elections as a case study. 

Partially dominating sets composed by influential profiles from different groups may reveal an internal organization of these communities and their ability to perform coordinated communication with the users. 
By comparing the reach of partially dominating sets, we may evaluate their ability to dominate the groups and segments of the entire network during some specific time window. 
We found that in the whole network the left-leaning high-influential information spreaders dominated, reaching a substantial fraction of the network with few spreaders. 
However, inside the groups, the results are inverse, and right-leaning high-influential users dominate their communities more easily (with fewer spreaders) than in the other analyzed groups. 


Our work also presents a methodology called $d-$modularity used for evaluating the contribution of specific groups to network's polarization by using the well-known modularity measure. 
We found that, on average, 44.1\% of the daily modularity of the interaction network was explained by right-leaning group's contribution, while the contribution of left-leaning group was only 13.4\%. 
This result is specially interesting since the right-leaning group is significantly smaller than the group of left-leaning users. 
The contribution of the right-leaning group to network's modularity is higher than the contribution of the left-leaning, significantly larger, group. 

These results bring evidence of extreme isolation and the ease of accomplishing coordinated communication that characterized right-leaning groups during the 2022 Brazilian elections. 
Based on the presented evidence, we may hypothesize that the isolation and the ease of domination inside this community influenced the coup events that occurred on 8 January 2023 in Brasília, which included the invasion of the Supreme Federal Court, the National Congress building, and the Planalto Presidential Mansion. 


\section*{Acknowledgments}

Ruben Interian was supported by research grant 2021/12456-5, São Paulo Research Foundation (FAPESP). 
Francisco Rodrigues acknowledges CNPq (grant 309266/2019-0) and FAPESP (grant 19/23293-0) for the financial support given for this research. 
Research carried out using the computational resources of the Center for Mathematical Sciences Applied to Industry (CeMEAI) funded by FAPESP (grant 2013/07375-0). 

\bibliographystyle{plain}
\bibliography{biblio}

\begin{thebibliography}{10}

\bibitem{Blondel2008}
Vincent~D. Blondel, Jean-Loup Guillaume, Renaud Lambiotte, and Etienne
  Lefebvre.
\newblock Fast unfolding of communities in large networks.
\newblock {\em Journal of Statistical Mechanics: Theory and Experiment},
  2008(10), 2008.

\bibitem{2015Campan}
Alina Campan, Traian Marius~Truta, and Matthew Beckerich.
\newblock Fast dominating set algorithms for social networks.
\newblock In {\em Proceedings of the 26th Modern AI and Cognitive Science
  Conference}, pages 55--62, 2015.

\bibitem{2004Chlebik}
Miroslav Chleb{\'i}k and Janka Chleb{\'i}kov{\'a}.
\newblock Approximation hardness of dominating set problems.
\newblock In Susanne Albers and Tomasz Radzik, editors, {\em Algorithms -- ESA
  2004}, pages 192--203, Berlin, Heidelberg, 2004. Springer Berlin Heidelberg.

\bibitem{Conover2011}
Michael~D. Conover, Bruno Goncalves, Jacob Ratkiewicz, Alessandro Flammini, and
  Filippo Menczer.
\newblock Predicting the political alignment of twitter users.
\newblock In {\em 2011 IEEE Third International Conference on Privacy,
  Security, Risk and Trust and 2011 IEEE Third International Conference on
  Social Computing}, pages 192--199, 2011.

\bibitem{2019Cota}
{Cota, Wesley}, {Ferreira, Silvio C.}, {Pastor-Satorras, Romualdo}, and
  {Starnini, Michele}.
\newblock Quantifying echo chamber effects in information spreading over
  political communication networks.
\newblock {\em EPJ Data Science}, 8(1):35, 2019.

\bibitem{2022case}
Casimiro Miguel~Vieira da~Silva~Ferreira.
\newblock Most influential tweet during the 2022 {B}razilian elections.
\newblock https://twitter.com/Casimiro/status/1584293297841111040, 2022.

\bibitem{2015DalMaso}
C.~{Dal Maso}, G.~Pompa, M.~Puliga, G.~Riotta, and A.~Chessa.
\newblock Voting behavior, coalitions and government strength through a complex
  network analysis.
\newblock {\em PLOS ONE}, 9:1--13, 2015.

\bibitem{2022pew}
D.~DeSilver.
\newblock {The polarization in today's Congress has roots that go back
  decades}.
\newblock Technical report, Pew Research Center, 2022.

\bibitem{2021Dinkelberg}
A.~Dinkelberg, C.~O'Reilly, P.~MacCarron, P.~J. Maher, and M.~Quayle.
\newblock Multidimensional polarization dynamics in us election data in the
  long term (2012–2020) and in the 2020 election cycle.
\newblock {\em Analyses of Social Issues and Public Policy}, 21:284--311, 2021.

\bibitem{2015Garcia}
D.~Garcia, A.~Abisheva, S.~Schweighofer, U.~Serdült, and F.~Schweitzer.
\newblock Ideological and temporal components of network polarization in online
  political participatory media.
\newblock {\em Policy and Internet}, 7:46--79, 2015.

\bibitem{2017Garimella_Long_term}
K.~Garimella and I.~Weber.
\newblock A long-term analysis of polarization on {T}witter.
\newblock In {\em Proceedings of the Eleventh International AAAI Conference on
  Web and Social Media}, pages 528--531, 2017.

\bibitem{2015Habibulla}
Yusupjan Habibulla, Jin-Hua Zhao, and Hai-Jun Zhou.
\newblock The directed dominating set problem: Generalized leaf removal and
  belief propagation.
\newblock In Jianxin Wang and Chee Yap, editors, {\em Frontiers in
  Algorithmics}, pages 78--88, Cham, 2015. Springer International Publishing.

\bibitem{2020Haq}
E.~ul Haq, T.~Braud, Y.~D. Kwon, and P.~Hui.
\newblock Enemy at the gate: Evolution of {Twitter} user's polarization during
  national crisis.
\newblock In {\em 2020 IEEE/ACM International Conference on Advances in Social
  Networks Analysis and Mining}, pages 212--216, 2020.

\bibitem{1998Haynes}
T.W. Haynes, S.~Hedetniemi, and P.~Slater.
\newblock {\em Fundamentals of Domination in Graphs}.
\newblock Taylor \& Francis, 1998.

\bibitem{2023wef}
S.~Heading and S.~Zahidi.
\newblock {The Global Risks Report 2023, 18th Edition}.
\newblock Technical report, World Economic Forum, 2023.

\bibitem{2022InterianRev}
R.~Interian, R.~G.~Marzo, I.~Mendoza, and C.~C. Ribeiro.
\newblock Network polarization, filter bubbles, and echo chambers: an annotated
  review of measures and reduction methods.
\newblock {\em International Transactions in Operational Research}, 2022.

\bibitem{2021IntMorRib}
R.~Interian, J.~R. Moreno, and C.~C. Ribeiro.
\newblock Polarization reduction by minimum-cardinality edge additions:
  {C}omplexity and integer programming approaches.
\newblock {\em International Transactions in Operational Research},
  28:1242--1264, 2021.

\bibitem{2018Interian}
R.~Interian and C.~C. Ribeiro.
\newblock An empirical investigation of network polarization.
\newblock {\em Applied Mathematics and Computation}, 339:651--662, 2018.

\bibitem{2015Kaleel}
Shakira~Banu Kaleel and Abdolreza Abhari.
\newblock Cluster-discovery of {T}witter messages for event detection and
  trending.
\newblock {\em Journal of Computational Science}, 6:47--57, 2015.

\bibitem{1990Kearns}
Michael~J. Kearns.
\newblock {\em The Computational Complexity of Machine Learning}.
\newblock ACM Distinguished Dissertation Award. Harvard University, 1990.

\bibitem{2010maoz}
Z.~Maoz and Z.~Somer-Topcu.
\newblock Political polarization and cabinet stability in multiparty systems:
  {A} social networks analysis of european parliaments, 1945-98.
\newblock {\em British Journal of Political Science}, 40:805--833, 2010.

\bibitem{2019Markgraf}
M.~Markgraf and M.~Schoch.
\newblock Quantification of echo chambers: {A} methodological framework
  considering multi-party systems.
\newblock In {\em 27th European Conference on Information Systems}, 2019.
\newblock Online reference at https://aisel.aisnet.org/ecis2019\_rp/91, last
  access on May 22, 2022.

\bibitem{2021RamaciottiMorales}
P.~R. Morales and J.-P. Cointet.
\newblock Auditing the effect of social network recommendations on polarization
  in geometrical ideological spaces.
\newblock In {\em Fifteenth ACM Conference on Recommender Systems}, pages
  627--632. ACM, 2021.

\bibitem{2016Newman}
M.~E.~J. Newman.
\newblock Equivalence between modularity optimization and maximum likelihood
  methods for community detection.
\newblock {\em Phys. Rev. E}, 94:052315, 2016.

\bibitem{2004Newman}
M.~E.~J. Newman and M.~Girvan.
\newblock Finding and evaluating community structure in networks.
\newblock {\em Phys. Rev. E}, 69:026113, 2004.

\bibitem{1991Parekh}
Abhay~K. Parekh.
\newblock Analysis of a greedy heuristic for finding small dominating sets in
  graphs.
\newblock {\em Information Processing Letters}, 39(5):237--240, 1991.

\bibitem{2017Shi}
Y.~Shi, K.~Mast, I.~Weber, A.~Kellum, and M.~Macy.
\newblock Cultural fault lines and political polarization.
\newblock In {\em Proceedings of the 2017 ACM Web Science Conference}, pages
  213--217. ACM, 2017.

\bibitem{Soares2019}
Felipe~Bonow Soares, Raquel Recuero, and Gabriela Zago.
\newblock Asymmetric polarization on {T}witter and the 2018 {B}razilian
  presidential elections.
\newblock In {\em Proceedings of the 10th International Conference on Social
  Media and Society}, SMSociety '19, page 67–76, New York, NY, USA, 2019.
  Association for Computing Machinery.

\bibitem{2022Weron}
Tomasz Weron and Katarzyna Sznajd-Weron.
\newblock On reaching the consensus by disagreeing.
\newblock {\em Journal of Computational Science}, 61:101667, 2022.

\bibitem{2021Wolfowicz}
M.~Wolfowicz, D.~Weisburd, and Hasisi B.
\newblock Examining the interactive effects of the filter bubble and the echo
  chamber on radicalization.
\newblock {\em Journal of Experimental Criminology}, 2021.
\newblock doi:10.1007/s11292-021-09471-0.

\end{thebibliography}

\end{document}